\documentclass[review]{elsarticle}

\usepackage{lineno,hyperref}
\usepackage{algorithm}
\usepackage{algpseudocode}
\usepackage{pifont}
\usepackage{booktabs}
\usepackage{graphicx}

\journal{Comp Phys Comm}

\def\b0{{\bf 0}}

\def\dg{\frac{d}{dg}}

\def\noi{\noindent}

\begin{document}

\begin{frontmatter}

\title{Hierarchical Parallelisation of Functional Renormalisation Group Calculations - hp-fRG}

\author{Daniel Rohe}
\address{J\"ulich Supercomputing Centre \\ Forschungszentrum J\"ulich GmbH \\ Email: d.rohe@fz-juelich.de}

\begin{abstract}
The functional renormalisation group (fRG) has evolved into a versatile tool in condensed matter theory for studying important aspects of correlated electron systems. Practical applications of the method often involve a high numerical effort, motivating the question in how far High Performance Computing (HPC) can leverage the approach.\\
In this work we report on a multi-level parallelisation of the underlying computational machinery and show that this can speed up the code by several orders of magnitude. This in turn can extend the applicability of the method to otherwise inaccessible cases. \\
We exploit three levels of parallelisation: Distributed computing by means of Message Passing (MPI), shared-memory computing using OpenMP, and vectorisation by means of SIMD units (single-instruction-multiple-data). Results are provided for two distinct High Performance Computing (HPC) platforms, namely the IBM-based BlueGene/Q system JUQUEEN and an Intel Sandy-Bridge-based development cluster. We discuss how certain issues and obstacles were overcome in the course of adapting the code. Most importantly, we conclude that this vast improvement can actually be accomplished by introducing only moderate changes to the code, such that this strategy may serve as a guideline for other researcher to likewise improve the efficiency of their codes.  
\end{abstract}


\end{frontmatter}


\section{Introduction}

Parallel algorithms are an essential ingredient when making proper use of high-performance computing (HPC) infrastructures. The parallelism of the underlying hardware has increased particularly fast during the last decade, and more importantly, has entered the realm of small to medium-sized computing infrastructures, extending even to the hardware that serves our every-day use. The habitual increase in cpu frequency has factually halted and Moore's law is only kept reasonably valid via an increase of parallel on-chip execution units. Therefore, concerning scientific applications we cannot rely anymore on an "automatic speed-up" by means of an increase in cpu frequency. Instead, it is rather the transfer of substantial, and sometimes even revolutionary, advances in HPC into the branches of fundamental science which in the near future can leverage and accelerate theoretical and experimental investigations which would otherwise remain unreachable. Harvesting the potential of parallel computing systems has never been a straight-forward task, with HPC being a scientific branch and community of its own for that very reason. 
We here report on an example of such a transfer and demonstrate that current HPC architectures can prove extremely helpful in accelerating a specific type of numerical application in quantum condensed matter physics, namely functional renormalisation group calculations. We stress from the very beginning that the concepts and the route taken to arrive at an enormous speed-up were not particularly complicated, and the actual code modifications turn out to be comparatively moderate and simple.  An all-important lesson we learned from this example is that we can avoid a complicated and intense code porting phase, something which scientific code developers hardly ever wish to engage in. 

\subsection{fRG} 

In recent times, functional renormalisation group methods have been established in quantum condensed matter physics as a powerful tool to detect and treat various
phenomena which arise in correlated electron systems. In particular, they can be used as an unbiased detector for competing correlations and quite a range of results is available for various models, parameter sets and specifically targeted observables \cite{HalbothMetzner:2000} \cite{Honerkamp:2001} \cite{HeddenMedenPruschkeSchonhammer:2004} \cite{HonerkampRoheAndergassenEnss:2004} \cite{KataninKampf:2004} \cite{KarraschHeddenPetersPruschkeSchonhammerMeden:2008} \cite{HusemannSalmhofer:2009} \cite{ReutherWolfle:2010} \cite{ReutherThomale:2011} \cite{GieringSalmhofer:2012} \cite{GerschHonerkampMetzner:2008}. A pioneering work in this context is given in \cite{ZanchiSchulz:1996}, for a comprehensive review see \cite{MetznerSalmhoferHonerkampMedenSchonhammer:2012}. The generic equation which governs the method is a functional differential equation which arises when an explicit scale-dependence is introduced in functionals of interest and a derivative with respect to this scale is taken subsequently. Upon expanding the functional of interest in terms of correlation functions, one obtains an infinite hierarchy of ordinary differential equation for these functions (fRG-ODE). The right-hand side (rhs) of this equation is in general given by multi-dimensional integrals and/or summations over weighted products of these functions, which stem from standard diagrammatic expansions in quantum field theory, see e.g. \cite{Berges:2002}.\\
In practice, the formal representation of these functions and equations is discretised and subsequently treated numerically. The resulting algorithms and codes can become very demanding in their computational intensity and thereby inhibitory to the overall scientific potential of the method, which in principle extends far beyond current applications. In this work we shall tackle this very obstacle by means of HPC methods, namely we wish to illustrate how a parallelisation at different levels can speed up a typical code of the fRG family such that ultimately a speed-up of several orders of magnitude is accomplished.\\
We feel that this step forward can be of considerable importance for one reason in particular: In very recent times an increasing amount of attention in computational science is paid to "virtual scanning techniques" in the area of materials design, the general idea of which is to pre-select potential candidates for new materials by first calculating relevant properties via numerical methods. We believe that the fRG has the potential to become a very helpful computational kernel within such strategies.
However, proper scanning of "theoretical materials" relies on the capability to generate results for a large variety of parameter sets in a short amount of time and by this virtue such quantitative improvements are a necessary ingredient for qualitative progress.\\

That said, the resulting motivation for this study is to provide a proof-of-concept that fRG calculations can massively profit from small and large-scale multi-parallel HPC environments. We thereby intend to offer a blueprint to enable fRG developers to best benefit from such architectures.\\

\subsection{HPC systems}

The key aspects when parallelising the numerical treatment of the fRG-ODE are imposed by a hierarchy of available means for parallelisation which is common to many contemporary HPC systems. In a bottom-up sequence this hierarchy can essentially be summarised as follows:\footnote{GPU-based accelerators are not considered in this work but are left for future studies. Yet, they are also highly parallel in nature and numerous aspects are likely to carry over to this case.}

\begin{itemize}
\item on-core single-instruction-multiple-data (SIMD) vector units, with a tendency of increasing width 
\item possibility/necessity to execute two or more processes/threads per core 
\item multiple cores per CPU, in part massively increasing in number (e.g. Intel Xeon Phi)
\item multiple CPUs per compute node with access to the same physical memory 
\item an ever increasing number of compute nodes connected by faster and faster networks
\end{itemize}

As an in-house example for a standard small-sized cluster we will refer to a general purpose Intel-based development cluster at JSC. As a representative system at the Petascale level we will refer to JUQUEEN, the IBM BG/Q supercomputer at JSC which offers $28,\!672$ nodes and $458,\!752$ cores in total \cite{StephanDocter:2015}. Each core can be overloaded by four processes such that ultimately $1,\!835,\!008$ concurrent threads can be executed. The set-up is summarised in Table \ref{hardware_set-up}.\\

 \begin{table}[tph]
\caption{HPC systems under consideration in this work}
\label{hardware_set-up}\centering
\begin{tabular}{l|l|r|rcc}
\toprule
Level & parallelisation type & Intel Cluster & JUQUEEN &\\ \midrule
distributive via &max number of threads & 512 & 1,835,008  &  \\
message passing &max nodes available& 16 &  28,672 &   \\ \midrule
shared memory &CPU/Node & 2 &  1 &   \\
via OpenMP & cores/CPU & 8 & 16 &  \\ 
 &max threads/core & 2  & 4  & \\
\midrule
vectorisation & SIMD & 256 Bit  & 256 Bit&                 \\
\bottomrule

\end{tabular}
\end{table}

In principle, the fRG-ODE constitutes an ideal candidate for parallelisation due to its high number of independent computational sub-tasks which are each subject to a high numerical effort. Yet, the goal of optimally harvesting all levels of parallelisation offered by the underlying hardware is somewhat subtle in certain aspects.
We will in the following report on a hierarchical parallelisation of a specific instance of an fRG code and comment on the experiences we encountered in the course of that process.

\section{Algorithm and Data Structure}

As mentioned above, in its exact formulation the functional renormalisation group equation leads to an infinite hierarchy of ordinary differential equations for correlation functions. In practical applications, this hierarchy is truncated and the functions of interest are discretised. This allows to map it onto a numerical algorithm and results in the computational task of solving an ODE with a large number of components and the need to plug the discretised functions into a multi-dimensional numerical quadrature/summation to obtain the rhs. The scale-dependent functions of interest are in most cases two-point (one-particle) and/or four-point (two-particle) correlation functions, also termed self-energies and effective interactions respectively. By storing the discretised representations of all these functions in a single array $V[i]$ of size $n$ where $i \in \{1,n\}$ and introducing $g$ as a continuous flow parameter, the fRGE has the form 

\begin{equation}
\dg V[i](g) = F[V[j](g);g]
\end{equation}  

\noi with initial condition $V[i](0) = V[i]_0$ and the functional $F[V[j]]$, $j \in \{1,n\}$, being non-linear and highly ''non-local'' regardless of choosing a real-space or reciprocal space representation of the problem. In this work we have restricted ourselves to the flow of the four-point function. It is straight forward to extend the strategy to other cases, e.g. to calculations of self-energies.\\
We use an explicit ODE stepper taken from a recent version of the Odeint library \cite{AhnertMulansky:2001}, which in the meantime has been added to the Boost C++ library collection \cite{Boost}. We use an explicit solver such that the computation of the rhs  can be done independently for each single component, since $V[j](g)$ is fixed for all $j$ when calculating  $F[V[j](g);g]$ within the algorithm.

Historically, the very first numerical computations based on an fRGE were effectuated on nearly completely non-parallel hardware, i.e. typically a single workstation using a single core, possibly including a narrow SIMD unit. This allows for a reasonable treatment of numerous aspects of interest even today. However, many aspects are not accessible within such a concept since computation times virtually explode when increasing either the resolution, or the dimensionality of the models of interest, or the desired accuracy of the results, etc. To account for this, in some cases the possibility of parallelising the algorithm has been exploited, e.g. in \cite{KarraschHeddenPetersPruschkeSchonhammerMeden:2008,KennesMeden:2013,RoheMetzner:2005}. While lacking detailed knowledge on all fRG codes in use we shall assume that such a first level of parallelisation has been implemented in other works as well.

We will account for the hierarchical hardware structure in a top-to-bottom fashion by organising the parallelisation in the following hierarchy of three levels:

\begin{itemize}
\item{distribution layer (via MPI): work packages are distributed to separate compute nodes which then act independently on \emph{physically separate} data}
\item{many-core/shared memory layer (via OpenMP): computation on a single node is distributed over all available physical/logical cores which refer to the same \emph{shared} data}
\item{vectorisation (via compiler): lowest level computational tasks are mapped to algorithmic variants which allow for an efficient use of vector units (SIMD)}
\end{itemize}

The paradigm we applied to each of these layers was to keep things as simple as possible while making optimal use of the respective potentials. We did \emph{not} aim at finding the \emph{best possible} hybrid algorithm, but we wanted to demonstrate that the possibilities of parallel architectures can, for this type of code, actually be exploited quite efficiently in simple ways. As stated above, we feel that this is an essential criterion when considering and/or proposing the migration of scientific applications to HPC infrastructures. In turn, this approach leaves room for further improvement which can be explored by more sophisticated means.

\section{Parallelisation}

We are dealing with an ODE for which the most time-consuming part consists in the calculation of the rhs. We therefore decided to employ a standard master-worker concept in which the workers are solely involved in calculating the rhs and the master executes the actual ODE step once the rhs has been calculated for all components. This setup explicitly introduces a serial element in the overall algorithm which needs to be watched closely since by Amdahl's law \cite{Amdahl:1967} it imposes an implicit limit for the strong scaling behaviour.\\ 
Moreover, if we eventually try to treat very fine discretisations, higher spatial or inner dimensions, or 
other schemes in which the number of components of $V[j]$ increases rapidly, we will eventually also run into memory capacity issues since the full array is replicated on all nodes. This is however beyond the scope of this work. A more general discussion along with conceptual details on efficient hybrid implementations of master-worker parallelism is given e.g. in \cite{Castellanos:2013}.

\subsection{Top-level parallelism: distributed computing via MPI} 

Keeping these \emph{a priori} limits in mind we chose a set-up for the calculation of the rhs sketched in Algorithm \ref{fRG_mpi_static} for a static work schedule and Algorithm \ref{fRG_mpi_dynamic} for a dynamic work schedule. Both variants will be discussed in the results section. We denote by 'node' the largest element of a (homogeneous) large-scale computing cluster which provides one common, shared memory to all its processing units, i.e. in most clusters a compute node of NUMA type (non-uniform memory access). 

\begin{algorithm}[H]
\caption{distribution layer - static scheduling}
\label{fRG_mpi_static}
master part
\begin{algorithmic}[1]
\Procedure{distribute $V[i]$ }{}
\State broadcast $V[i]$ to each node $n$ \Pisymbol{psy}{206} $N$ 
\EndProcedure
\Procedure{collect results }{}
\For{each node $n$ \Pisymbol{psy}{206} $N$ }
\State receive from each node the corresponding results
\EndFor
\EndProcedure
\end{algorithmic}
worker part
\begin{algorithmic}[1]
\Procedure{receive $V[i]$ }{}
\State receive broadcast of $V[i]$ from master
\EndProcedure
\Procedure{do calculation}{}
\For{each index to calculate}
\State calculate and locally store result
\EndFor
\EndProcedure
\Procedure{send results }{}
\State send results to master
\EndProcedure
\end{algorithmic}
\end{algorithm}

\begin{algorithm}[H]
\caption{distribution layer - dynamic scheduling}
\label{fRG_mpi_dynamic}
master part
\begin{algorithmic}[1]
\Procedure{distribute $V[i]$ }{}
\State broadcast $V[i]$ to each node $n$ \Pisymbol{psy}{206} $N$ 
\EndProcedure
\Procedure{distribute single work units and collect results}{}
\State send single work units to workers
\While{work units left to calculate}
\State collect single results from any worker and assign next unit
\EndWhile
\State collect remaining results
\EndProcedure
\end{algorithmic}
worker part
\begin{algorithmic}[1]
\Procedure{do calculation}{}
\While{work to be done}
\State receive work unit from master
\State compute result for single unit
\State send result to master
\EndWhile
\EndProcedure
\end{algorithmic}
\end{algorithm}

In practice we use the Message Passing Interface standard (MPI) to implement these common algorithms. We also anticipate to use only one MPI rank per node in order to reduce the limits imposed by memory size per node. We could relax this condition by allowing for several ranks per node which would generalise the algorithm accordingly. We shall see later that for a small and intermediate number of nodes dynamic scheduling is preferable, while for a very high number of nodes the static concept is required for proper scaling. 

\subsection{Second-level parallelism: multi-core/shared memory layer}

Once all values of $V[i](g)$ at a given value of $g$ have been distributed to the workers, the numerical task on a single compute node consists in general in a multi-dimensional numerical quadrature (summation) over a continuous (discrete) space for each of the components assigned to this node. In the case looked at here this amounts to a two-dimensional integration over the two-dimensional Brillouin zone of the Hubbard model on a square lattice \cite{HonerkampRoheAndergassenEnss:2004}. While there exist very efficient serial algorithms for numerical quadrature, there are little to no resources available when it comes to generically efficient parallel quadrature algorithms, in particular if the amount of cores used in a shared memory environment increases to very high numbers. Therefore we have chosen a fixed grid mid-point rule for the outer integral subject to quadrature, since this can be parallelised very efficiently by means of OpenMP. For the inner integral we used a locally adaptive mid-point rule which then constitutes the largest inseparable numerical block to be executed by a single OpenMP task. The fact that we apply the distribution of work on the outer quadrature is intended to ensure that the computational time needed for each core to calculate the function value is reasonably long compared to the OpenMP overhead. The validity of this assumption needs to be verified by appropriate analysis tools as well as scaling tests as will be described in later sections. \\
This is again a rather simple way of introducing parallelism which of course leaves room for improvement by itself. Yet, we will see that it is an appropriate way to arrive at a very efficient performance level.

\subsection{Third-level parallelisation: vectorisation and core overload}

In order to exploit the advantages of vectorisation as well as other cpu-internal means of parallelism we follow a similar strategy as in the previous section, however from a bottom-up view. We wish to avoid the usage of intrinsics or even assembly-like programming in order to keep the code easily portable between different HPC systems as well as easily sustainable and maintainable. Modern compilers are able to vectorise code if it is offered to them in a sufficiently simple and obvious fashion, which is typically a single loop with a fixed iteration count, be it at compile time or at runtime, within which simple mathematical operations are applied to elements residing in elementary, contiguous arrays. We thus split the \emph{innermost} quadrature kernel over multiple domains and evaluate these domains by means of such loops. To verify the efficiency of this procedure we need to ensure that the compiler does indeed vectorise the relevant code sections and we have to check by scaling benchmarks and, if possible, hardware-oriented tools whether we arrive at the desired speed-up. \\
In addition to this compiler-generated auto-vectorisation, current architectures support and often recommend the execution of multiple threads per core to make proper use of the complex internal architecture. We will include this core overloading in our analysis in combination with the aspects of vectorisation.
We learned from the two distinct combinations of HPC architecture and compiler, that vectorisation and core overloading seem to be intimately related. On the X86 platform it is a combination of both which yields a reasonable gain at single core level, while on JUQUEEN the compiler avoids explicit vectorisation and rather resorts to intrinsic ways to make use of the underlying hardware parallelism. This seemed somewhat disappointing at first sight, considering the behaviour on the X86 side. However, it turned out that the final structure of the code was just as beneficial for core performance on the BG/Q. Again, it is possible and promising to dig much deeper into technical details, preferably at the level of assembly instructions, but here we wish to keep things as simple as possible while obtaining a substantial gain.

\section{Scaling results on small clusters}
 
As our working example we focus on the interaction flow method introduced in  \cite{HonerkampRoheAndergassenEnss:2004,PolonyiSailer:2002}, applied to the two-dimensional Hubbard model on a square lattice with nearest-neighbour hopping amplitude $t$ set to unity, next-nearest neighbour hopping $t' = -0.01t$, temperature $T=0.1t$ and chemical potential $\mu=-0.02$. This corresponds to a hole-like case with the bare Fermi energy being slightly above the van-Hove energy and the Fermi surface being closed around $(\pi,\pi)$. The bare on-site interaction is directly dependent on $g$ and serves as the flow parameter.  
We started from the code base used in \cite{RoheMetzner:2005} which we adapted to the interaction flow and subsequently altered step by step in order to make use of \emph{all} the levels of parallelisation as outlined in the previous section. We will restrict ourselves to results for the average time needed to calculate the rhs within the first ODE step, for which seven single, complete evaluations of the rhs are required. This is sufficient for the analysis of the scaling behaviour.\\ 
The calculation of the rhs consists in a numerical treatment of three distinct one-loop diagrams. It amounts to a quadrature over the two-dimensional Brillouin zone, which is numerically split into an outer integral over an energy variable and an inner integral over a radial variable. The outer variable is pinned to a fixed, equidistant grid of 1600 points, giving an energy spacing of $0.005t$. The inner quadrature is treated by a custom adaptive routine with a relative target precision of $10^{-4}$ and a maximum of $2187$ points at which the innermost loop-kernel is evaluated.\\
 
In the following sections we analyse the efficiency of the individual parallelisation layers in a bottom-up fashion, ending up with scaling results for a fully hybrid code running on up to 16 compute nodes. We used a typical set of model parameters with 32 patches and a discretisation in analogy to \cite{HonerkampRoheAndergassenEnss:2004} and \cite{RoheMetzner:2005}. This leads to an fRGE for $4072$ independent coupling components when all spacial, spin-related and time-reversal symmetries are taken advantage of. \\
All results were obtained on a JSC-internal compute cluster based on Intel E5-2650 CPUs running at 2.0 GHz. We used the Intel C++ compiler 2015 (neglecting minor updates) and the ParaStation MPI implementation version 5.1.2 by ParTec, unless stated otherwise. In addition we offer some results at the single node level obtained on the Intel Xeon Phi architecture (KNC) for comparison.
 
 \subsection{Single-core: baseline, vectorisation and core-overload}
 
We begin with computations on a single core in order to determine an appropriate reference value to compare against when we later check the scaling behaviour. For this purpose we ran the code on a single thread and switched off vectorisation explicitly. The first row of Table \ref{single_core} shows an average time for computing the rhs of $1271.71$ seconds of the non-vectorised code when executed using one thread and defines this as the initial baseline.
 
Next we exploit the lowest level of parallelism by using the SIMD processing units, in this case one 256 Bit AVX unit per core. As mentioned above, the code was structured in a way to carry out elementary operations on simple arrays of double-precision floating point numbers in loops with pre-determined iteration counts, and the respective arrays were explicitly aligned. This facilitates the compilers built-in ability to vectorise code efficiently. The level of vectorisation was checked in compiler reports and/or by checking the generated assembly code.\\

Furthermore, current architectures are designed to offer potential performance gains when dispatching several threads on a single core. In our case the effect of this core overload offers a similar performance gain compared with vectorisation, for the architectures under consideration here. However, already in some immediate follow-up cpu architectures the SIMD width has been doubled, such that the benefit of - and thus need for - vectorisation in this respect will increase. 

We see from Table \ref{single_core} that it is indeed favourable to employ both, vectorisation and an overloaded core in order to arrive at a speed-up factor of $2$ compared to the baseline result. It should be mentioned that the ideal speed-up one can expect from mere vectorisation is of course $4$, if the code at the lowest level consisted exclusively of elementary mathematical operations which are executed in a four-fold parallel fashion compared to the serial case. However, mainly two aspects  reduce the true gap between vectorised and non-vectorised code. First, the vectorisation process as such introduces an overhead at the lowest hardware level. This is particularly disadvantageous for small iteration counts, which we deal with here. Second, there are some complex operations executed before and after the vectorised loop which are unavoidable and reduce the maximum gain from speeding up the parallelised regions. While this is a small draw-back on the X86 cluster, we will see later that these effects are more prominent on other hardware.\\ 
By deeper substantial architectural software changes, we could most likely reach better scaling on a single core. Yet, we managed to stick to small and simple code changes and gained a factor of two while staying on a single compute core, which is by all means satisfactory at this level. 
 
 \begin{table}[tph]
\caption{Single core parallelism}
\label{single_core}\centering
\begin{tabular}{clrrc}
\toprule
Threads & vectorised & rhs avg [s] & speed-up \\ 
              &                  &  &  \\ \midrule
1 & no  & 1271.71 & 1                \\\midrule
1 & yes & 889.30 &  1.43 \\
2 & no  & 1021.16  &   1.25  \\
2 & yes &  628.41 & 2.02 \\\bottomrule
\end{tabular}
\end{table} 

 \subsection{Adding intra-node parallelism using OpenMP}

As our new baseline we now use the shortest runtime of $628.41$ seconds of the vectorised code when using two threads on a single core and introduce shared-memory parallelism as described in the previous section. Table \ref{omp_scaling} again shows the average times for computing the rhs along with the corresponding speed-up values. All results are obtained with a compact affinity, meaning that a physical core is always occupied by two threads in order to extract the correct scaling with respect to the number of physical cores.

 \begin{table}[tph]
\caption{OpenMP parallelism}
\label{omp_scaling}\centering
\begin{tabular}{clrrr}
\toprule
cores  & threads & rhs avg [s] & speed-up \\ 
               &   &  &  \\ \midrule
1 & 2 & 628.41 & 1                \\\midrule
2 &  4 & 320.60 & 1.96 \\
4 &  8 &  175.62  & 3.58 \\
8 &  16 &   91.79 & 6.85 \\
16 &  32 &  46.87  &  13.41\\\bottomrule
\end{tabular}
\end{table} 

We observe that we manage to gain a very reasonable speed-up factor of $13.41$ when moving from one to $16$ cores, which is quite satisfactory. We thus managed to efficiently parallelise the code at this level by means of slight code changes. 

 \subsection{Adding distributed-memory parallelism using Message Passing Interface (MPI)}
 
Finally we show in Table \ref{mpi_scaling} the results when the work is distributed over up to 16 compute nodes using a dynamic MPI strategy as sketched in Algorithm  \ref{fRG_mpi_dynamic}. The fastest implementation of $32$ threads per node from the previous section is used as the new baseline, with the slight difference between $46.87$ and $47.5$ seconds being due to the intrinsic variation in runtime between different runs. 
 
\begin{table}[tph]
\caption{MPI parallelism}
\label{mpi_scaling}\centering
\begin{tabular}{clccc}
\toprule
worker nodes  & rhs avg [s] & speed-up \\ \midrule
1 & 47.15  & 1                \\\midrule
2 & 23.65 &   1.99  \\
4 & 11.82 &   3.99  \\
8 &  5.91 &   7.98 \\
16 &  2.97 &  15.88      \\\bottomrule
\end{tabular}
\end{table} 

The code scales nearly perfectly, showing a linear speed-up in the number of worker nodes. We chose to exclude the master node from any computational task for reasons of scalability. Of course, for such a small number of nodes it can be reasonable to also defer some computational work to the master, the disadvantage being that any delay on the master node with respect to the workers translates into a larger and larger waste of cpu time the more nodes we use. Furthermore	, for very high numbers of worker nodes the master will have to cope with a high load due to MPI communication as we will see later. This will be another motivation to keep it free from computational tasks.\\

Gathering the results from all three levels of parallelisation in Table \ref{total_small_cluster}, we find that we managed to reduce the average time to compute the rhs from $20$ minutes to $3$ seconds. That is, we arrive at a total speed-up of about  $430$ when comparing the non-vectorised code run on a single thread and a single core to the case of using $512$ threads on $256$ cores.

 \begin{table}[tph]
\caption{total parallelism}
\label{total_small_cluster}\centering
\begin{tabular}{lllrrc}
\toprule
cores & threads & vectorised & rhs avg [s] & speed-up \\ 
             &              &    &  &  \\ \midrule
1     & 1 &no  & 1271.71 & 1                \\\midrule
256 & 512 &yes &  2.97 & 428 \\\bottomrule
\end{tabular}
\end{table} 
 
 \subsection{Pushing OpenMP to the limits: Scaling on Intel Xeon Phi (KNC)}
 
It is instructive to benchmark the code on hardware which offers an even higher level of parallelism on the shared-memory level in order to push towards the limits of node-internal scalability. For this purpose we ran it on a single instance of Intels Xeon Phi (Knight's Corner) running at 1.0 GHz and offering 60 cores, each with a SIMD width of 512 Bit, i.e. eight double precision numbers.\\
Additional platforms which are worth to be examined further are GPU accelerators, however this is outside the scope of this report. We note that this option was used in \cite{KennesMeden:2013}.
 
 \subsubsection{Vectorisation}
 
We again start with a non-vectorised executable which we run on a single core and we end up with an average time of $10,\!959$ seconds, which is roughly nine times as long as the counterpart run on a standard core as described in the previous sections. This is not too surprising taking into account the major architectural differences between the two cpu architectures.
  
A priori we could expect that the use of vectorisation offers about double as much speed-up as in the case of a 256 Bit vector unit used in the X86 cluster nodes. However, as shown in Table \ref{mic_scaling_vectorisation} the speed-up of $1.8$ accomplished by mere vectorisation seems rather low. This corroborates the conjectures from the previous section concerning the limitations of vectorisation, and it is actually known that on this very architecture the need and benefit of properly prepared inner loops with preferably large iteration counts is crucial to exploit the potential of the 512 Bit SIMD units. To put it differently, when using this platform we truly exceed the limit of on-core scalability \emph{for this version of the code}. The necessity for an algorithm to exploit vectorisation on Intel's Xeon Phi is outlined in detail in \cite{Jeffers:2013}. We could now try to push things further by altering the algorithm to make better use of these hardware capabilities, and this is indeed scheduled for future work. In fact, some procedure like this is nearly always applied when the underlying hardware structure changes. Here, we only wish to illustrate one kind of limit that we ran into in the process of arriving at a highly scalable code. (We note that there is no such thing as a highly-scalable code as such, but there are rather pairs of the type (code,hardware) which can result in a highly-scalable \emph{execution set}.)\\

\begin{table}[tph]
\caption{Intel Xeon Phi vectorisation}
\label{mic_scaling_vectorisation}\centering
\begin{tabular}{clccc}
\toprule
Threads  & vectorised & rhs avg [s] & speed-up \\ 
               &   &  &  \\ \midrule
1 &  no   &  10959.71 & 1                \\
1 &  yes &  5992.57 &  1.83               \\\midrule
\end{tabular}
\end{table}

\subsubsection{OpenMP}
 
The Xeon Phi accelerator used here (5110P) offers 60 physical cores for computational purposes. However, at the internal level each core runs two independent threads which are served at every second clock tick, roughly speaking. Therefore, it will be imperative to trigger the OpenMP implementation to put at least two threads on each core. On top of this we may overload each of these two states as we did on the standard nodes, and in fact the Xeon Phi reports 240 logical cpus in total.

Table \ref{mic_scaling_omp} summarises the OpenMP scaling results with the single-core vectorised runtime as the baseline. The picture resembles the behaviour on standard nodes: When using the maximum number of logical threads we can essentially exploit the physical limit of scalability and arrive at a maximum speed-up of $116.5$ when using $240$ OpenMP threads.
 
  \begin{table}[tph]
\caption{Intel Xeon Phi OpenMP parallelism}
\label{mic_scaling_omp}\centering
\begin{tabular}{clccc}
\toprule
Threads  &  & rhs avg [s] & speed-up \\ 
               &   &  &  \\ \midrule
1 &  &  5992.57 & 1                \\\midrule
60 &  &  102.57 &  58.4  \\
120 &  &   65.71 &  91.2 \\
240 &  &   51.43 & 116.5 \\\bottomrule
\end{tabular}
\end{table}
 
Ultimately, we observe that the final runtime is within ten percent of the runtime on a standard node, cf. Table \ref{omp_scaling},  which means we could already switch to such an infrastructure without relevant disadvantages in the scientific workflow. Also, in terms of energy consumption according to hardware specs the accelerator requires about as much power as two E5 2650 CPUs, such that by this measure the accelerator already constitutes a valid alternative even without additional architecture-specific code optimisations.\footnote{cf. specs available at http://ark.intel.com} 
However, considering that nominally the peak performance of the Xeon Phi accelerator is about four times as high compared to a dual-cpu node of the X86 cluster, a more thorough adaptation of the code could and should lead to substantial further optimisations on this class of hardware.
 
\subsection{Wrap-up: Single nodes and small clusters}

While the scaling results presented above for the fRG specific task seem trivially plausible, straight forward and not far from ideal, this is by no means a generic property of any fRG code. The route taken to finally arrive at such a scaling behaviour did involve several changes in code structure and design, often due to issues in the areas of load balancing and compiler-related optimisation. 

The most time consuming part, however, consisted in \emph{locating} the bottlenecks and from this deducing the code changes. The \emph{resulting} code is kept reasonably simple and the actual modifications were relatively few. At the same time, the speed-up which can be accomplished even for such a small number of compute nodes is more than considerable. We note that we started with an average time to calculate the rhs of about 1200 seconds, i.e. 20 minutes. Building a complete fRG run on this set-up would imply a total runtime of the order of a few days, depending of course also on the desired accuracy, the parameter sets used and other conditions. The hybrid code when deployed on 16 nodes allows for step times of less than a minute for one ODE step, implying durations for a complete run of the order of less than two hours. In terms of future applicability of fRG methods for material scans already this small-scale parallelisation can thus be considered a potential can opener, since scanning involves a high number of runs applied to many different parameter sets and model configurations. 

\section{Scaling on large clusters}

We have seen in the previous sections that already compute clusters of moderate size can substantially speed up fRG calculations and reduce the runtime by more than two orders of magnitude. Next, we will investigate in how far this strategy carries over to much larger systems, be it in order to reach a further speed-up, which in HPC language is termed "strong" scaling, or be it to treat (much) larger vector sizes, which in HPC language is termed "weak" scaling.

\subsection{Results on IBM BlueGene/Q (JUQUEEN)}

JUQUEEN is an IBM BlueGene/Q HPC infrastructure, operated at and by J\"ulich Supercomputing Centre (JSC) at Forschungszentrum J\"ulich, in cooperation with IBM. Inaugurated in 2012 and operational at full capacity since the beginning of 2013, it remains one of the most powerful supercomputers worldwide, listed on rank eleven in the TOP 500 supercomputer ranking of November 2015. It consists of 28 racks, $28,\!672$ nodes and a total of $458,\!752$ cores \cite{StephanDocter:2015}, where each core provides a $256$ Bit SIMD/vector unit and permits a four-fold thread overload.\\
In Table \ref{mpi_scaling_juqueen_1} we show scaling results starting from a baseline of $32$ nodes, i.e. $32 \times 16 = 512$ compute cores. This starting point approximately corresponds to the case of $16$ nodes on the X86 cluster, in terms of runtime as well as power demand.\\ 
In addition, we will now also vary the problem size to motivate an intrinsic scaling limit. With an ODE vector of size $4072$ and a structure which relies on the computation of  the rhs for one single component on one node, it is clear \emph{a priori} that strong scaling will saturate at least when the number of reaches the order of the vector size. Yet, going from $32$ nodes to $1024$ nodes, i.e. by an upscaling of $32$, we note that the runtime needed for one single ODE step still decreases substantially and the strong scaling behaviour remains very reasonable even up to an average load of four components per node. This is due to the use of dynamic MPI scheduling, i.e. the fact that we use Algorithm \ref{fRG_mpi_dynamic}. The slightly super-linear speed-up is due to the fact that the master node does not take part in the computation and for technical reasons we here work with the total number of nodes rather than the number of worker nodes, unlike in the case of the X86 cluster. In the limit of a large number of nodes this is irrelevant for practical purposes.\\
 
\begin{table}[tph]
\caption{Strong scaling for a small-sized problem on JUQUEEN (IBM BG/Q)}
\label{mpi_scaling_juqueen_1}\centering
\begin{tabular}{lr|rc}
\toprule
Size: & 4072  & schedule: & dynamic\\ \midrule
node  & nodes & rhs avg & speed-up \\
 cards&              &  [s] &   \\ \midrule
 1 & 32      &  3.92 & 1                \\\midrule
 2 & 64      &  1.93 & 2.03 \\  
 4 & 128    &  0.97 &  4,04  \\
 8 & 256    &   0.50 &   7.84 \\
 16 & 512   &  0.26 &  15.07 \\
 32 & 1024  &   0.15 &  26.13 \\
 64 & 2048  &   0.09 &  43 \\ \bottomrule
\end{tabular}
\end{table}

Next, we move to an intermediate problem size for which the coupling vector consists of $110,\!136$ independent components. In Table \ref{mpi_scaling_juqueen_2} we again include results for Algorithm  \ref{fRG_mpi_dynamic}, i.e. a fully dynamic MPI scheduling in which each worker is assigned the task of computing the rhs for a single component, returns the result, and receives the next work unit. This concept is however known to fail eventually in an HPC context of massively parallel distributive set-ups, since the master node will congest with respect to communication and the resulting MPI overhead spoils the scaling. Indeed we observe that scaling starts to breaks down when moving from $8,\!192$ to $16,\!384$ nodes.
A simple recipe to overcome this consists in changing from a dynamic MPI work load distribution to a static one by assigning a certain set of predefined work units to every worker task first, have the workers compute all results, and finally collect all results in packages. Such a static assignment has of course disadvantages, one being the lack of implicit load balancing. E.g. the clustering of computationally more demanding tasks to one or a few nodes could spoil the scaling behaviour even more than the dynamic version. To prevent this we employed a strided distribution of the static workload. This turned out to be fully sufficient to achieve very good scaling up to the full machine size of $28,\!672$ nodes. 

We observe for such a static assignment following Algorithm  \ref{fRG_mpi_static}, which we have further supplemented by a strided distribution of the workload to avoid clustering of heavy load at a single worker, that this is sufficient to achieve very good scaling up to the full machine size of $28,\!672$ nodes. 

\begin{figure}[ht]
\caption{Scaling behaviour for a medium-sized problem on JUQUEEN. A plot for the small-sized problem is included for comparison.}
\label{scaling_juqueen}
\includegraphics[width=\linewidth]{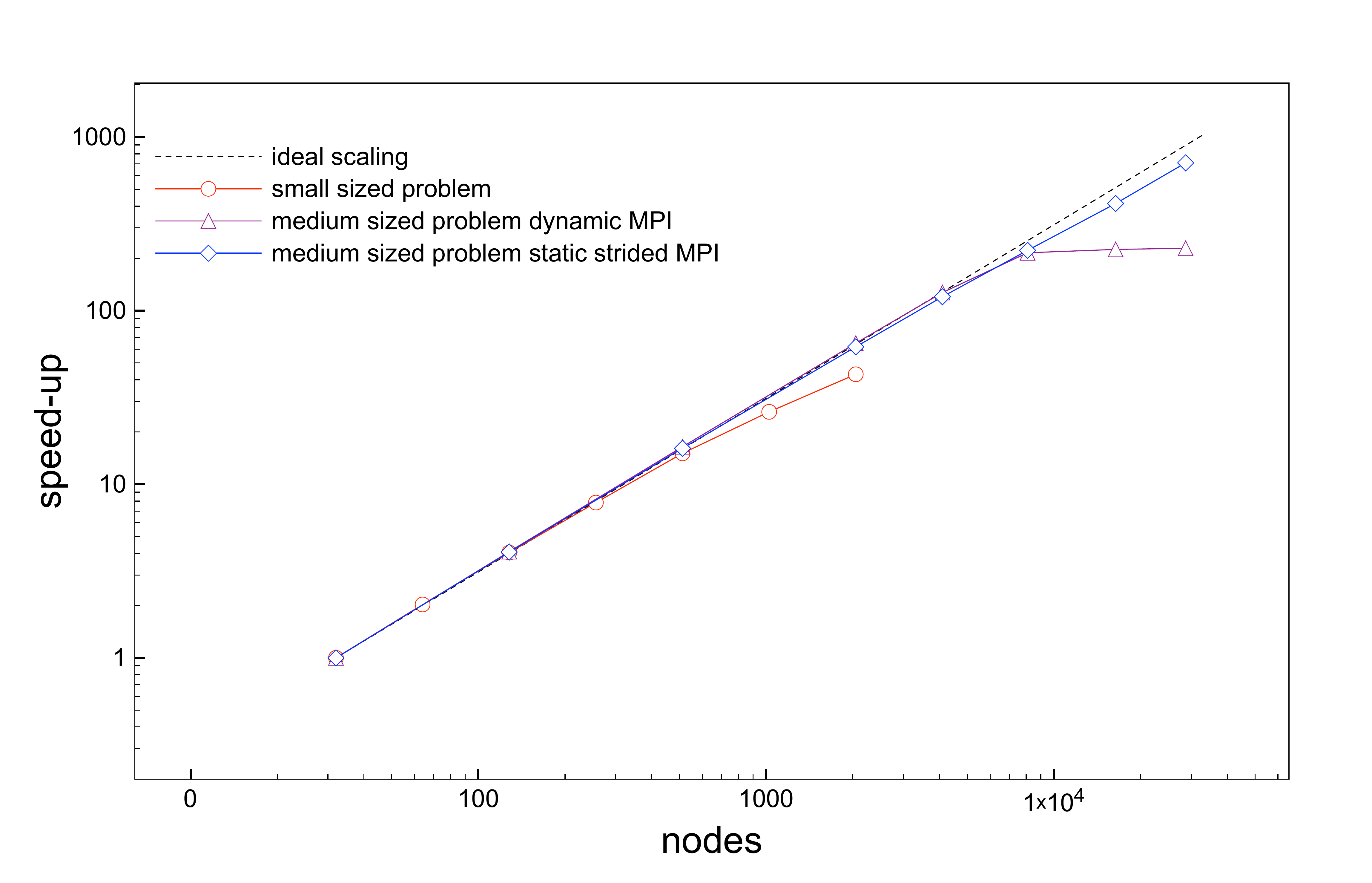}
\end{figure}

Figure \ref{scaling_juqueen} graphically summarises the scaling bevaviour and illustrates that the static, strided work load distribution scales up to the whole machine size of JUQUEEN. \\

While scaling is an important and central "observable" in any HPC context, it is essential from the scientists point of view to ask the question if there is a qualitative gain one may achieve by this scaling. In that respect what counts are actual runtimes and treatable system sizes and we note that in the limit of a very large number of nodes it is possible to squeeze the time needed to calculate the rhs down to about a second for a moderate system size. For what matters to the fRG community this is the all important outcome since it opens the door even further towards the possibility of materials scanning. 

\begin{table}[tph]
\caption{Strong scaling for a medium-sized problem on JUQUEEN (IBM BG/Q)}
\label{mpi_scaling_juqueen_2}\centering
\begin{tabular}{lr|rc|rc}
\toprule
Size: & 110136  & schedule: & dynamic & schedule: & static (strided) \\ \midrule
node  & nodes & rhs avg & speed-up   & rhs avg & speed-up   \\
 cards&              &  [s] &   &  [s] & \\ \midrule
 1 & 32      &   148.69 & 1    & 149.1 &  1           \\\midrule
 4 & 128     &   36.34 & 4.09 & 36.64 & 4.07\\  
 16 & 512    &   9.04 &  16.44 & 9.25 & 16.12 \\
 64 & 2048    &   2.29 &   64.93 & 2.41 & 61.87 \\
 128 & 4096   &   1.17 &  127.08 & 1.24 & 120.24 \\
 256 & 8192  &  0.69 &  215,49 & 0.67 & 222.53 \\
 512 & 16384  &  0.66 & 225.29 & 0.36 & 414.17 \\ 
 896 & 28672  &  0.65 & 228.75& 0.21 & 710.0 \\ \bottomrule
\end{tabular}
\end{table} 

\section{Issues, obstacles and tools}

As already mentioned, the general set-up of the fRG-ODE constitutes a scenario which seems ideal for very efficient and at the same time simple parallelisation techniques over a wide range of relevant parameters. We started with an initial code that had originally been optimised for serial performance and which scaled perfectly only for a small number of MPI ranks. The outcome is a code which now benefits from two additional levels of parallelisation at the shared-memory and SIMD/core overload level and scales to very large numbers of nodes at the distributed memory level. The evolution involved various means of redesign and restructuring, including changes in the low-lying algorithms and continuous evaluations of the appropriate locations where parallelisation-specific pragmas and functions had to be placed. Some examples of such an approach along with generic issues which are also valid in our context are e.g. given in \cite{Jin:2011}, a discussion along with general conceptual details on efficient hybrid implementations of master-worker parallelism is given in \cite{Castellanos:2013}. \\
The final code which scales over several orders of magnitude is then again rather simple, which was a core statement already in the introduction. We thus hope that the obstacles we had to overcome on the way to this simple code structure may motivate others to engage in a similar strategy and to help making the transition to a scalable codes an easier task than it may seem.\\

An obligatory step when encountering a break-down in scaling is to analyse the application with respect to its usage of the underlying hardware infrastructure. This is possible thanks to a number of HPC-specific tools which, when used in combination,  allow for correctness checking and performance analysis at nearly all technical and algorithmic levels. In this section we will outline the two most relevant issues we encountered together with a brief description of how these issues were analysed and which tools were of great use in doing so.
  
\subsection{Making use of vectorisation and core overload}

The initial code neither involved shared-memory multi-threading, nor did it offer simple means for a compiler to make efficient use of on-core vector units and other parallelism. The first step was to start from the very bottom of the algorithm, which essentially means at the very lowest level of basic evaluations of computational kernels within a multi-dimensional numerical quadrature. For that purpose we needed to ensure that loops run over a minimum of iterations for vectorisation to become reasonably effective, and we wanted to end up with code that makes it sufficiently easy for compilers to automatically vectorise these loops without forcing us to resort to intrinsics. While it can be argued that this approach is possibly less efficient than optimisation "by hand" we wanted to keep the code generically portable. 

An inevitable subtask when trying to find and evaluate means for efficient vectorisation are harware-based tools that shed sufficient light on the inner workings of the actual binary. In case of the Intel-based X86 cluster we mainly relied on Intel's native performance analysis tools VTune and Intel Advisor, which provide information on various aspects and potential bottlenecks. In particular, on KNC hardware it allows for an explicit analysis of the actual vectorisation intensity which is achieved in reality. In combination with compiler diagnostics offered by the Intel compiler, we were able to combine the information obtained from the tools and the reported vectorisation level of the compiler to efficiently analyse bottlenecks and potential locations for code changes. This eventually allowed us to alter the code as much as needed but no more than necessary to obtain a high level of vectorisation intensity.\\

Porting and tuning the code on JUQUEEN relied mainly on the Score-P/Scalasca/Vampir toolset. Here, it actually turned out that the compiler decides to not explicitly vectorise the code but rather to optimise it for on-core parallelism as it can be used when overloading a core. Yet, it proved to be an efficient algorithm also in this case and showed that the path we were guided to on the X86 cluster was a valid direction also for lifting the code to a BG/Q environment.

While it was very valuable to enable compiler-based vectorisation, the gain at this level of parallelisation was a little lower than we anticipated on the standard X86 hardware. In case of the Xeon Phi it was much lower, and in the case of JUQUEEN it was even preferable for the compiler to resort to other means of optimisation. This tells us that there is still room for improvement, since in principle and also in ideal examples vectorisation can be much more efficient. Considering the current hardware evolution with wider and wider SIMD units appearing in standard cpus, there is a big potential in this area to further improve the code. When designing a similar code from scratch we would strongly suggest to address this at a very early stage of code development.
 
\subsection{Load balancing vs. logistical overhead}

When distributing work over a large number of nodes, already slight differences in work load can completely spoil the overall efficiency of parallelisation.  Therefore it is imperative to construct the overall algorithm in such a way that the load balancing amongst all nodes is extremely well equalised.

The same argument holds for the intra-node load balancing, where it applies to the work distribution among the available threads. This of course means that in summary we need to ensure a near-to-perfect load balancing between all available threads on all participating nodes, i.e. ultimately for full machine runs on JUQUEEN between $1,\!835,\!008$ threads ($28,\!672$ nodes $\times$ $16$ cores $\times$ $4$-fold thread overload). 

This requirement translates into the necessity to find the best location for both levels of parallelisation, MPI and OpenMP-based, in the numerical algorithm: If the level in the algorithm is too high up, we will spoil the load balancing since we cannot smoothen out differences in compute time. Therefore, from this point of view we would place both handles as low as possible while keeping sufficient space between them.

If we followed this paradigm we would, however, eventually run into difficulties at the other end of the scale. If we reduce the thread-specific numerical effort at the shared-memory OpenMP level by too much, the logistical overhead that is intrinsic to the OpenMP runtime framework will itself turn into a bottle-neck and act as a limiting factor for any further scaling when increasing the intensity of parallelisation. Similarly, if we reduce the individual work packages per node by too much, i.e. if each node is "too quick", it is the MPI communication overhead which will constitute a bottleneck in an analogous manner. 

These requirements impose a lower limit to the level in the algorithm where it is sensible to place the two parallelisation techniques MPI and OpenMP. We could even be unlucky and hit the case when there is no sweet-spot or sweet-region between the aforementioned approaches, and for very small problem sizes we indeed observed this when making use of the full JUQUEEN.

\begin{figure} 
\caption{Example for the load distribution for static MPI work load scheduling and outer level OpenMP when using 8 worker tasks each employing 32 OpenMP threads}
\label{static_mpi_outer_omp}
\includegraphics[width=\linewidth]{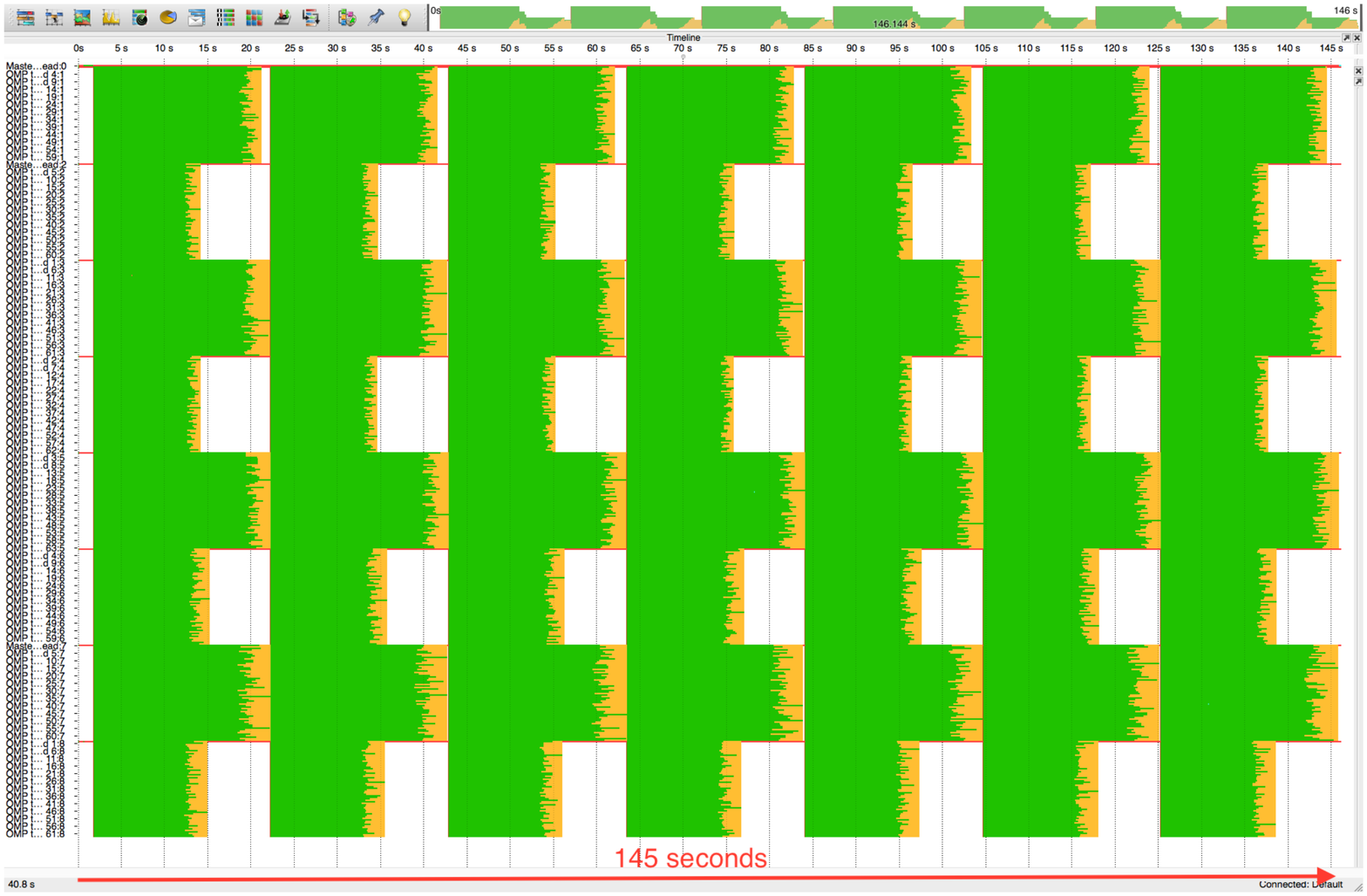}
\end{figure}

\begin{figure} 
\caption{As Fig. \ref{static_mpi_outer_omp} but for static MPI work load scheduling and inner-level OpenMP}
 \label{static_mpi_inner_omp}
\includegraphics[width=\linewidth]{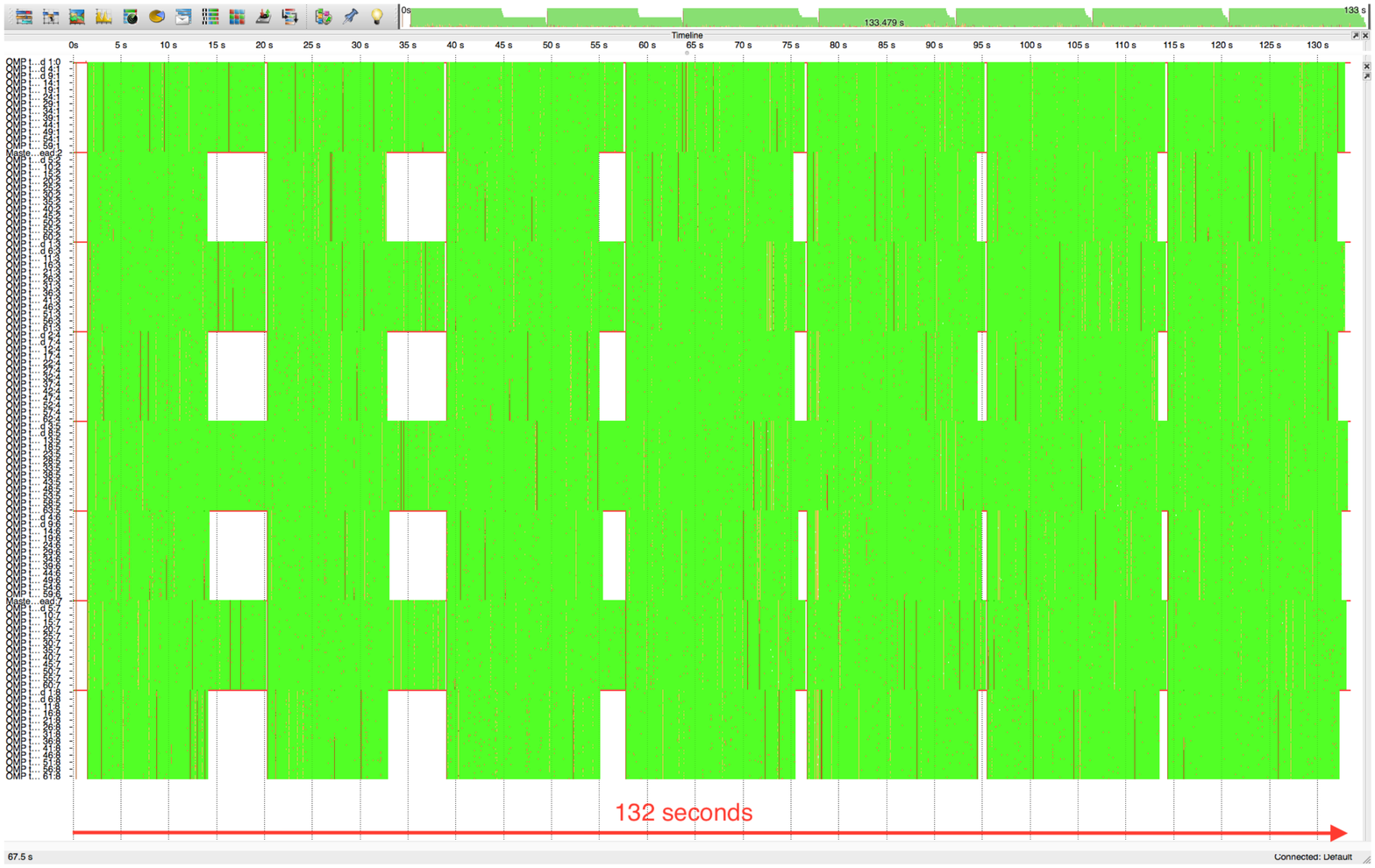}
\end{figure}

\begin{figure} 
\caption{As Fig. \ref{static_mpi_outer_omp} but for dynamic MPI work load scheduling and inner-level OpenMP}
 \label{dynamic_mpi_inner_omp}
\includegraphics[width=\linewidth]{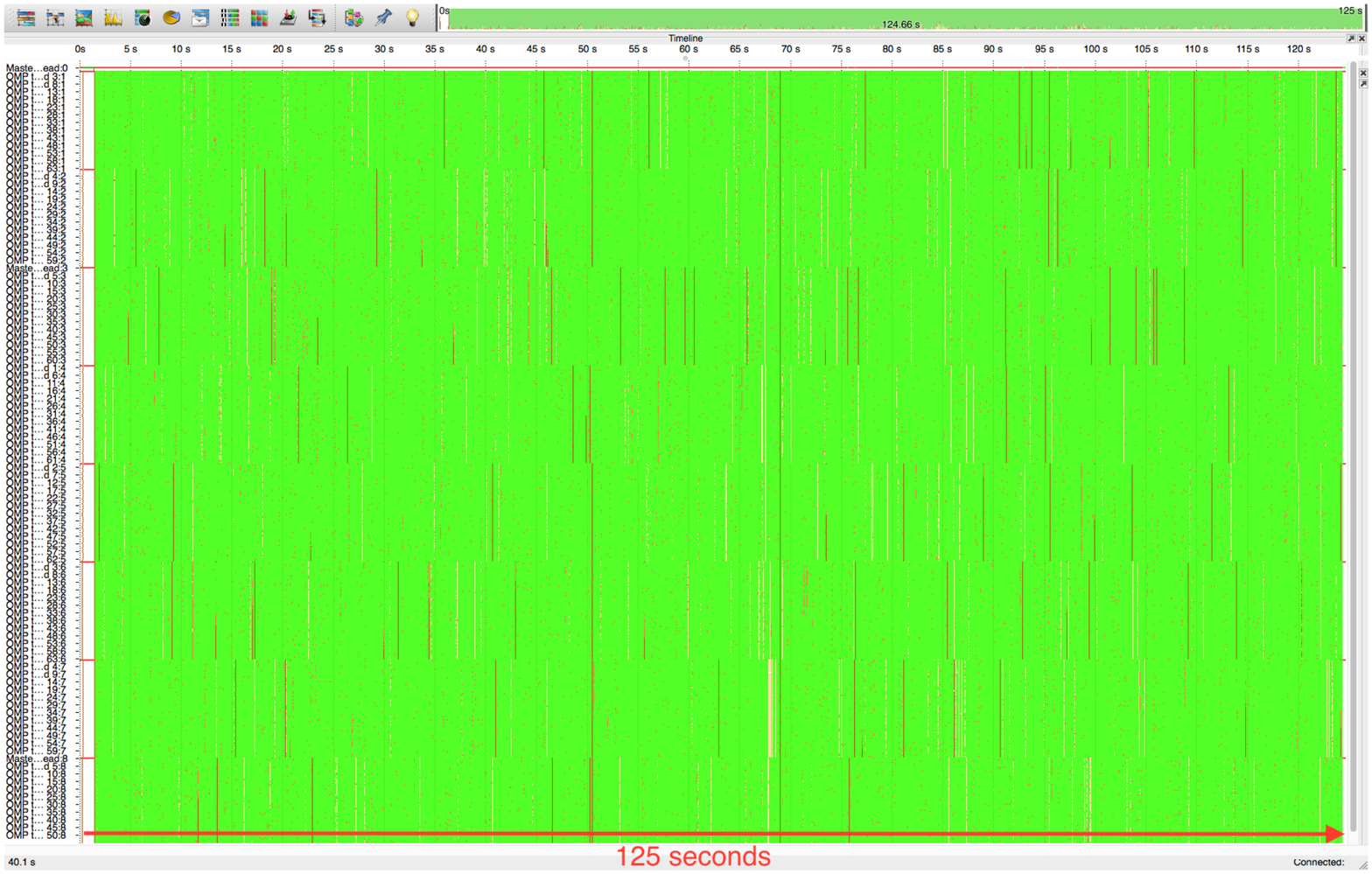}
\end{figure}

The process of locating such a sweet-spot would be a more or less impossible task without dedicated tools which have been designed and developed over many years for the specific purpose of analysing massively parallel code. In our case we mainly relied on the tools Score-P, Scalasca and Vampir \cite{Mohr:2014} for the analysis of MPI and OpenMP load balancing as well as overhead analysis. In order to illustrate a typical part of the workflow when analysing the quality of the load balancing we show three figures for the case of using eight MPI worker nodes, each of which employs $32$ OpenMP threads. Figure \ref{static_mpi_outer_omp} shows a sample taken via Score-P/Vampir which shows the load distribution for the case when each MPI rank is assigned a fixed set of components at the beginning of each rhs evaluation. The OpenMP distribution within an MPI rank is such that each OpenMP thread calculates the rhs for one single component. While this was appropriate for certain cases and a small number of nodes, the graphical analysis immediately unveals the draw-backs. The white regions show that half of the nodes run idle for a substantial period of time during each calculation of the rhs. The seven evaluations of the rhs necessary to execute one ODE step can clearly be seen.
This deficiency was first tackled by changing the work scheduling at the OpenMP level by moving the OpenMP parallelisation one level down. Namely, instead of having each thread compute the full rhs for a single component we make use of all threads already for the computation of the outer numerical quadrature in the two-dimensional integration, as described above. This leads to a load balancing picture shown in Figure \ref{static_mpi_inner_omp} where it can be seen that we obtain quite a reasonable level.
We can further improve the situation by switching to the dynamical master-worker concept of assigning only one component of the rhs at the time to each rank, receiving the result and reassigning the next task to this now available rank. As shown in Figure \ref{dynamic_mpi_inner_omp} this further increases the efficiency of the algorithm and leads to a nearly ideal load balancing. On the other hand, this strategy increases the intensity of communication between master and workers and is known not to scale beyond a certain number of nodes, depending on the specific application. Yet, we note that it is well worth considering and benchmarking this standard approach as we found that it constitutes the most efficient algorithm up to $4096$ nodes. Beyond this limit it is more efficient to resort to a static MPI load distribution.

\section{Conclusion}

The central purpose of this work - and more generally of HPC in science - is to provide means to numerically tackle scientific problems of interest which would otherwise be unfeasible or extremely time consuming. However, HPC systems cannot ''automagically'' speed up any scientific application. This is due to the fact that the enormous nominal peak performance relies on an extremely high level of hardware parallelism. Therefore, a key task to harvest the power of HPC consists in designing applications in a manner that they scale on such parallel infrastructures. In many cases this task appears to be a major obstacle to code developers, the result often being the implicit decision to ''maybe do it at a later stage''. \\
In case of a particular example of a functional renormalisation group (fRG) code we find that an HPC-ready design can indeed be accomplished, while at the same time the necessary code modifications can be kept at a very moderate level. In our case legacy code needed to be altered and the final resulting code is simple. Arriving at this result involved a number of intermediate steps of analysis and code redesign, which we also sketched in this paper.  \\
In summary, we hope to motivate the fRG community to use parallel computing architectures more intensely and eventually move to full-fledged HPC systems in order to most efficiently harvest the potential of the method. \\
As the closing bottom line we note that we achieved an overall speed-up of about five orders of magnitude, coming from a single core computation on an Intel-based compute node to a full run on a 28 rack BG/Q system. This is an enormous spread in compute time and in our eyes justifies the common conjecture that ''quantity is quality'' in this context.

\section{Acknowledgements}

It became clear at the very beginning of this project that the task of efficiently parallelising the code would need regular input from several sides with respect to specific details at several stages. The successful outcome demonstrates that a cooperation between experts from different scientific fields can greatly improve the efficiency of a particular code, and by this the efficiency of science.
Along this line, the author greatly acknowledges regular support and assistance by HPC experts at JSC, in particular Brian Wylie from the cross-sectional team "Performance Analysis", whose help was invaluable for many of the individual tasks of locating and removing "the next bottleneck". Equally helpful was the assistance accessible through the partnership with Intel, mainly offered by Heinrich Bockhorst and Zakhar Matveev, which leveraged the value of Intel tools and helped to identify many crucial aspects at the interface between code and actual hardware. In a similar manner we received input and advice on behalf of IBM/Lenovo through Christoph Pospiech.\\
Data acquisition and analysis were greatly facilitated by using the J\"ulich Benchmarking Environment (JUBE) \cite{Luehrs:2015}, and a special thanks goes to Sebastian L\"uhrs for implementing some very helpful features at occasional phases of the project.



\end{document}